\documentclass[10pt,twocolumn]{article}

\usepackage[T1]{fontenc}
\usepackage[utf8]{inputenc}
\usepackage{lmodern}

\usepackage[a4paper,margin=1.8cm]{geometry}

\usepackage{amsmath,amssymb,amsfonts}

\usepackage{graphicx}
\usepackage{float}

\usepackage[authoryear,round]{natbib}

\usepackage[colorlinks=true,
            linkcolor=blue,
            citecolor=cyan,
            urlcolor=blue]{hyperref}

\usepackage{caption}
\usepackage{subcaption}

\usepackage{titlesec}
\titleformat{\section}{\fontsize{13.2}{15}\bfseries}{\thesection}{0.5em}{}

\makeatletter
\renewenvironment{thebibliography}[1]
 {\section*{\refname}%
  \small
  \list{\@biblabel{\@arabic\c@enumiv}}%
       {\setlength{\labelwidth}{1.5em}%
        \setlength{\labelsep}{0.5em}%
        \setlength{\leftmargin}{3em}% margen total
        \setlength{\itemindent}{-2em}% hace el hanging indent
        \setlength{\itemsep}{2pt}%
        \setlength{\parsep}{0pt}%
        \setlength{\topsep}{0pt}}%
  \sloppy\clubpenalty4000\widowpenalty4000%
  \sfcode`\.\@m}
 {\endlist}
\makeatother

\begin{document}
\title{\textbf{Can Light Cross a Singularity? Exact Solutions from Analogue Gravity}}

\renewcommand{\thefootnote}{\fnsymbol{footnote}}
\author{Juan Manuel Paez$^{1}$\footnotemark[1], Franco Fiorini$^{2}$\footnotemark[2], Santiago M. Hernandez$^{2}$\footnotemark[3]}
\date{}

\twocolumn[
\maketitle

\begin{center}
{\footnotesize
$^{1}$Instituto de Astronomía y Física del Espacio (IAFE, CONICET-UBA), Ciudad Universitaria, Buenos Aires, Argentina \\
$^{2}$Grupo de Comunicaciones Ópticas, Departamento de Ingeniería en Telecomunicaciones, Consejo Nacional de Investigaciones Científicas y Técnicas (CONICET) and Instituto Balseiro (UNCUYO), Centro Atómico Bariloche, Av.~Ezequiel Bustillo 9500, CP8400, S. C. de Bariloche, Río Negro, Argentina
}
\end{center}

\vspace{0.1cm}

\begin{center}
\begin{minipage}{0.8\textwidth}
\small
\begin{center}
\textbf{Abstract}
\end{center}

Using a simple spacetime hosting a strong curvature naked singularity, we employ an analogue gravity model to study electromagnetic fields in this background. We find exact solutions to the full set of electrostatic and electrodynamic equations that remain regular even in the presence of the singularity. Moreover, certain solutions sustain a regular and bounded power flux across the singularity, suggesting that electromagnetic energy may be transmitted through it.

\end{minipage}
\end{center}

\vspace{0.8cm}
]
\footnotetext[1]{jmpaez@iafe.uba.ar}
\footnotetext[2]{franco.fiorini@ib.edu.ar}
\footnotetext[3]{shernandez@ib.edu.ar}

\setcounter{footnote}{0}
\renewcommand{\thefootnote}{\arabic{footnote}}

\section{Introduction}
Strong curvature singularities are among the most striking predictions of general relativity. Characterized by the divergence of curvature invariants and the breakdown of the classical description of spacetime, they are commonly associated with extreme and often destructive physical effects, such as the unbounded growth of tidal forces and the lack of predictability \citep[][]{HE, J1, Naber, Senovilla, J2}. In their vicinity, classical field theories are typically expected to lose validity, leading to divergent physical quantities and preventing the propagation of well-defined signals. Understanding whether any meaningful physical information can survive or propagate in the presence of spacetime singularities therefore remains a fundamental open problem \citep[][]{singularitycrossing1,singularitycrossing2}.

In this context, the study of electrodynamics in curved spacetime provides a natural starting point for exploring the interaction between classical electromagnetic (EM) fields and extreme gravitational environments \citep[e.g.,][]{Blandford, Thorne1, Thorne2}. The generally covariant formulation of Maxwell’s equations allows EM fields to be defined on arbitrary spacetime backgrounds, while explicitly encoding the influence of curvature on their propagation \citep[e.g.,][]{Cohen, Tsagas, Lobo}. A remarkable "accident", firstly pointed out independently by \cite{Gordon} and \cite{Tamm}, and later on rediscovered by \cite{Pleb}, establishes an analogy between the propagation of EM fields on a curved background spacetime and electrodynamics in flat spacetime filled with a material medium subjected to certain non-conventional constitutive relations. In this way, this formalism encodes the influence of spacetime geometry on the effective electromagnetic properties of the medium. This is the base of the (optical) \textit{analogue gravity} models \citep[][]{Analog1,Analog2}, as well as the realm of \textit{transformation optics} \citep[][]{Leon1,Leon2}.    

In this work, we focus on the electrodynamics on a toy-model spacetime containing a naked, strong curvature singularity. We exploit the exact character of the resulting equations to analyze the behavior of electromagnetic fields near the singularity and to explore the possibility of signal transmission across it. This contribution is based on \citet{maestria1,maestria2}. 

\section{Waves in Plebanski-Tamm Media}
Following the seminal ideas of \cite{Pleb} and \cite{Tamm}, the equations governing electromagnetism in an arbitrary curved spacetime can be written in the same form as those describing a material medium in flat space obeying the Plebanski-Tamm (PT) constitutive equations. More explicitly, the source-free Maxwell equations in a curved background are
\begin{equation}
    F^{\mu\nu}_{\;\;;\mu} = 0\,, \quad \quad \quad \quad (\epsilon^{\mu\nu\rho\sigma}F_{\rho\sigma})_{,\mu} = 0,
\end{equation}
where $F_{\mu\nu}=A_{\nu,\mu}- A_{\mu,\nu}$ and $A_\mu=(\phi,\bar{A})$ is the four-potential. On the other hand, for a material medium in 3D flat space, the electromagnetic equations read (in units where $c=1/\sqrt{\epsilon_0 \mu_0}=1$),
\begin{align}
    &\bar{\nabla} \cdot \bar{D} = 0\,, \quad \quad \bar{\nabla} \times \bar{E} + \frac{\partial \bar{B}}{\partial t} = 0,\,\label{eq:maxwell_medioscom} \\
    & \bar{\nabla} \cdot \bar{B} = 0\,, \quad \quad \bar{\nabla} \times \bar{H} - \frac{\partial \bar{D}}{\partial t} = 0\,.\notag
\end{align}
The analogy arises when Cartesian coordinates are used and the material medium also satisfies the PT constitutive equations
\begin{align}
    &\bar{D} = \mathbf{K}\bar{E} + \bar{\Gamma} \times \bar{H}, \label{pt_1}\\
    &\bar{B} = \mathbf{K}\bar{H} - \bar{\Gamma} \times \bar{E}, \label{pt_2}
\end{align}
where the Cartesian components of the matrix $\textbf{K}$ and the vector $\bar{\Gamma}$ are connected to the spacetime geometry as
\begin{equation}
    K_{ij}=-\sqrt{-g}\,g^{ij}/g_{tt}\,,\,\,\,\,\,\,\,\,
    \Gamma_m=g_{tm}/g_{tt}\,.\label{defmatk}
\end{equation}
In this way, the effects coming out from the study of EM fields on a given spacetime background $(M,g_{\mu \nu})$ can be viewed as EM phenomena in conventional 3D, flat Euclidean space, provided it is filled with a rather strange material medium whose constitutive relations are given by Eqs. (\ref{pt_1}) and (\ref{pt_2}). This fact was extensively utilized for modeling intrinsically high-curvature optical effects \citep[see e.g.][]{Mackay1, Mackay2, Oleg, Turner,Falcon}. More aligned with our present concerns, the PT analogy serves also as a tool to inquire on questions of more fundamental character associated to the structure of space-time, as the possibility of having causal anomalies \citep[][]{Barcelo}. Worth of mention is the fact that possible uncontrolled effects arising as the blowing up of some curvature scalars, are viewed in the PT formalism not as singularities in the space-time (which is perfectly regular because is flat space), but as pathologies in the structure of the matrix $\textbf{K}$, which plays a double role of permeability and permittivity in Eqs. (\ref{pt_1}) and (\ref{pt_2}).

In this work, we consider the EM waves
\begin{align}
    &\bar{\mathcal{E}}(\bar{x},t)=\bar{E}(\bar{x})\exp\big[i\,k_{0}\,(\bar{k}(\bar{x})\cdot\bar{x}- t)\big], \notag\\
    &\bar{\mathcal{H}}(\bar{x},t)=\bar{H}(\bar{x})\exp\big[i\,k_{0}\,( \bar{k}(\bar{x})\cdot\bar{x}- t)\big],\label{expEyH2}
\end{align}
where $\bar{k}(\bar{x})$ is the dimensionless wave vector and $k_0=\omega$ is the flat-space wave number, i.e. the usual wave number characterizing a monochromatic plane wave in empty flat space. This last case is obtained from (\ref{pt_1}) and (\ref{pt_2}) by taking $\textbf{K}=\textbf{I}$ and $\bar{\Gamma}=\bar{0}$, corresponding to Minkowski spacetime with metric $g_{\mu\nu}=diag(-1,1,1,1)$. Substituting the ansatz (\ref{expEyH2}) into Eqs. (\ref{eq:maxwell_medioscom}), and using (\ref{pt_1}) and (\ref{pt_2}), we obtain the following two equations coming from Faraday and Ampère-Maxwell laws 
\begin{align}
    &i\,k_{0}^{-1}\,\bar{\nabla}\times\bar{E}=[\bar{\nabla}(\bar{k}\cdot\bar{x})+\bar{\Gamma}]\times\bar{E}-\textbf{K} \bar{H}\,, \label{eq:rot_E_cuasi_planas}\\
    &i\,k_{0}^{-1}\,\bar{\nabla}\times\bar{H}=[\bar{\nabla}(\bar{k}\cdot\bar{x})+\bar{\Gamma}]\times\bar{H}+\textbf{K} \bar{E}\label{eq:rot_H_cuasi_planas}\,.
\end{align}
These two equations are sufficient to characterize the EM field, because the two divergence equations in (\ref{eq:maxwell_medioscom}) are not really independent when the ansatz  (\ref{expEyH2}) is considered; see  \cite{maestria1}. 

\section{A mild, naked singularity}

\begin{figure}
    \centering
    \includegraphics[width=0.8\linewidth]{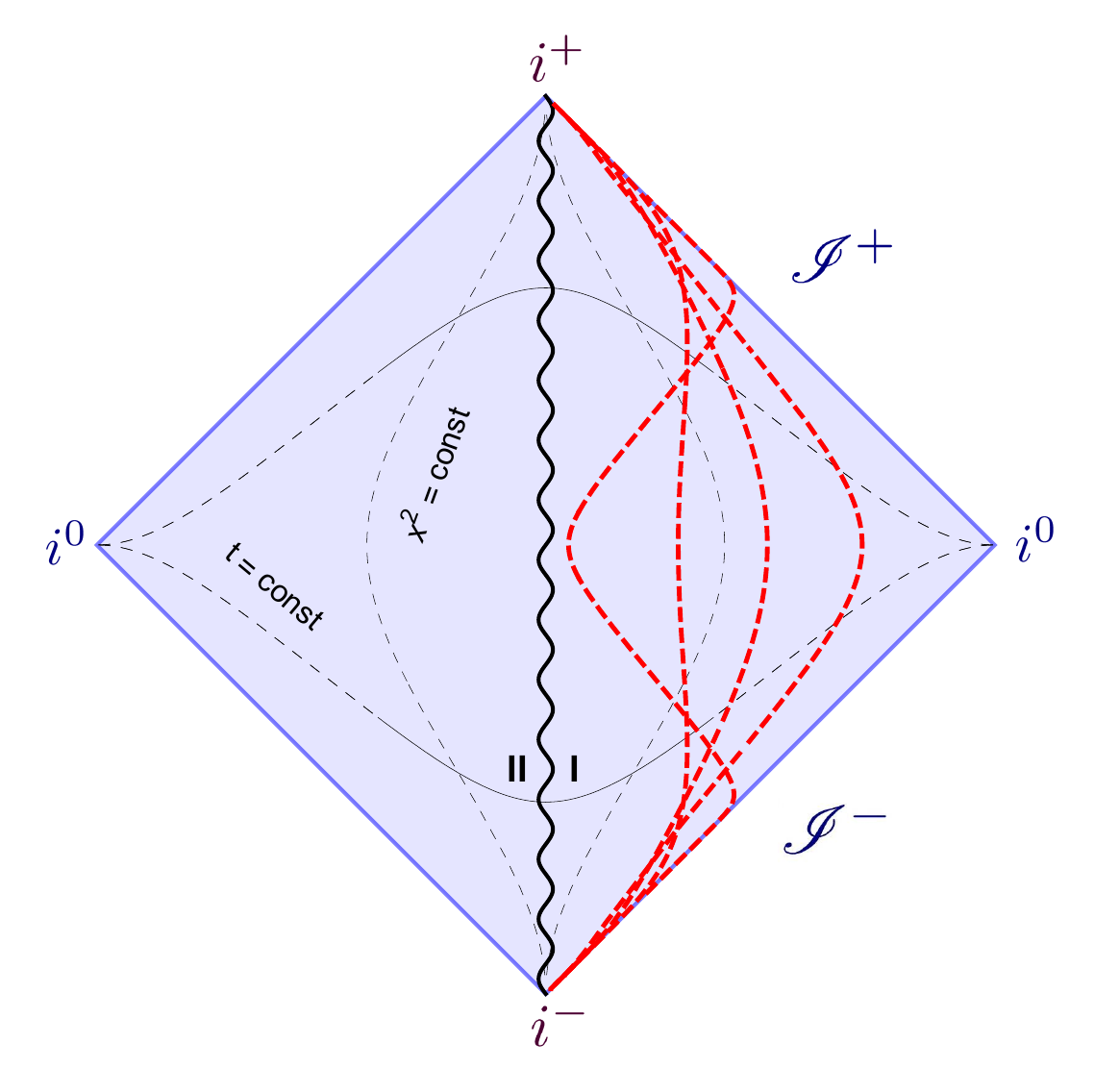}
    \caption{Global structure of the spacetime given by Eq. \ref{mettoy-1}. Some timelike geodesics are shown in red. The wavy black curve is the naked, timelike singularity at $x=0$.}
    \label{fig:penrose_diagram}
\end{figure}

With the aim of exploring the behavior of light in close proximity to a strong curvature, naked singularity, we have constructed a toy model spacetime $(\mathbb{R}^4, g_{\mu\nu})$ in which the line element is given by \citep[][]{maestria1}
\begin{equation}\label{mettoy-1}
    ds^2=-x^{-2}dt^2+dx^2+dy^2+dz^2\,, \end{equation}
which gives rise to only two non-null components of the Ricci tensor $R_{tt}=2 x^{-4},\,\,\,\, R_{xx}=-2 \,x^{-2}$. The space-time, whose global structure is encoded in Fig. \ref{fig:penrose_diagram}, has a curvature singularity at $x=0$, as witnessed by the blow up there of various invariants associated to the curvature, i.e.,  $R=g^{\mu\nu}R_{\mu\nu}=-4\,x^{-2}$,  $\mathcal{R}^2=\mathcal{K}=2R^2$ where $\mathcal{R}^2=R^{\mu\nu}R_{\mu\nu}$ and $\mathcal{K}=R_{\mu\nu\rho\sigma}R^{\mu\nu\rho\sigma}$. 
Due to the simplicity of the metric, we can fully characterize the null geodesics of the space, which correspond to trajectories of light rays as they come from geometrical optics (GO) in the analogue medium description. This approximation involves the limit $k_0\rightarrow \infty$ in the LHS of Eqs. (\ref{eq:rot_E_cuasi_planas}) and (\ref{eq:rot_H_cuasi_planas}), as well as taking $\bar{\nabla}(\bar{k}\cdot\bar{x})\approx\bar{k}$ on the RHS of them. The Hamilton motion equations describing the light rays come from the optical Hamiltonian \citep[see ][]{Hamop,Sluijter1,Sluijter2}
\begin{equation}\label{hamilton}
\mathrm{H}_{am}\doteq\det(\textbf{K}) -\bar{p}^{\,\intercal} \textbf{K}\,\bar{p}=0,
\end{equation}
and they are \cite[][]{nos}
\begin{eqnarray}
\frac{d \bar{x}}{dt}&=&-2 \textbf{K}\bar{p},\label{hamiltoncoorfin}\\
\frac{d \bar{k}}{dt}&=&\bar{p}^{\,\intercal}\Big[ [\textbf{K}_{i}-tr(\textbf{K}^{-1}\textbf{K}_{i})\,\textbf{K}]\,\bar{p}+2\textbf{K}\,\bar{p}_{i}\Big]\hat{e}_{i}\,,\label{hamiltonmomfin}
\end{eqnarray}
where $\bar{p}=\bar{k}+\bar{\Gamma}$, $\hat{e}_{i}$ are the Cartesian unit vectors in $\mathbb{R}^3$, $tr(\textbf{A})$ is the trace of the matrix $\textbf{A}$, and $\textbf{K}_{i}$ are three matrices whose components are obtained from the components of $\textbf{K}$ by differentiating with respect to the coordinate $x_{i}$. Notice that (\ref{hamilton}) is basically the dispersion relation.

\begin{figure}[h]
    \centering
    \includegraphics[width=0.8\linewidth]{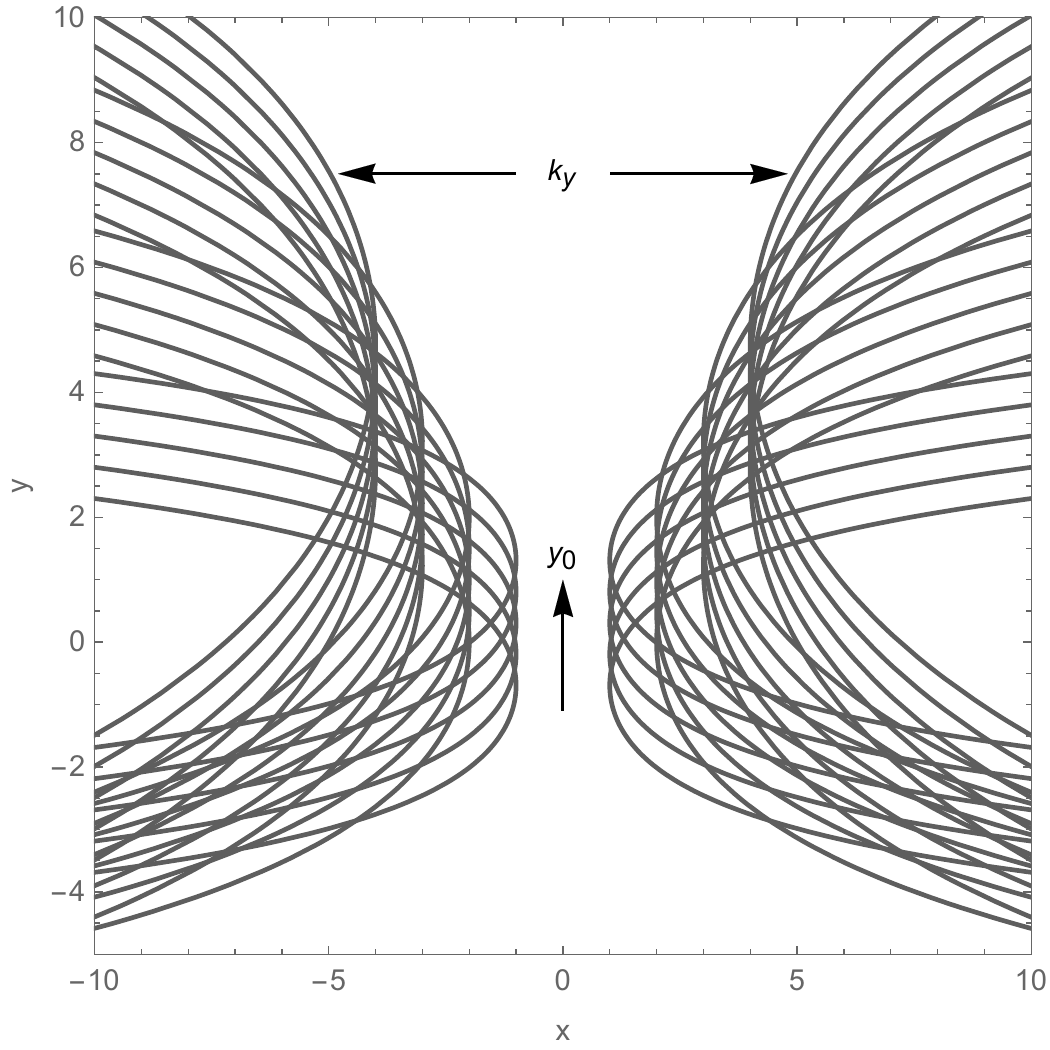}
    \caption{Trajectories of light in the $x-y$ plane according to GO; these are null geodesics of the spacetime.}
    \label{fig:geodesics}
\end{figure}

The definitions (\ref{defmatk}) applied to the metric (\ref{mettoy-1}) lead to 
\begin{equation}\label{eq:K}
    \textbf{K}=|x|\,\textbf{I}\,, \,\,\,\,\, \bar{\Gamma} =\bar{0}\,,
\end{equation}
so the dispersion relation  (\ref{hamilton}) turns out to be 
\begin{equation}
    \mathrm{H}_{am}=|x|\left(x^2-|\bar{k}|^2\right)=0\,,\label{ham}
\end{equation}
which implies $x^2=|\bar{k}|^2$. Sets of null geodesics with $k_z=0$, i.e., lying in the $x-y$ plane, as they come out from Eqs. (\ref{hamiltoncoorfin}) and (\ref{hamiltonmomfin}), are shown in Fig. \ref{fig:geodesics}. The evolution of the $x$ coordinate is described by the solutions
\begin{eqnarray}
    x(t)&=&\pm|\Tilde{k}| \csc{(2|\Tilde{k}|\,t+t_{0})}\,,\,\,\,\,\,\Tilde{k}\neq 0,
    \label{eq:sol_x}\\
x(t)&=&\pm(2\,t+t_{0})^{-1}\,,\,\,\,\,\,\tilde{k}=0,
\label{eq:sol_x_k_0}
\end{eqnarray}
and $y(t),z(t)$ read
\begin{equation}
\left\{\begin{array}{rl}
       y(t) \\
       z(t)\\
    \end{array}\right\}=\left\{\begin{array}{rl}
       y_{0} \\
       z_{0}\\
\end{array}\right\}\pm\left\{\begin{array}{rl}
       k_{y} \\
       k_{z}\\
\end{array}\right\}\log\left|\tan(|\Tilde{k}|\, t+t_{0}/2)\right|\,,  \label{eq:sol_y}
\end{equation}
where $t_{0}$, $y_{0}$ and $z_{0}$ are integration constants and $\Tilde{k}\doteq k_{y}^2+k_{z}^2$, thus, $\Tilde{k}=0$ represents rectilinear propagation orthogonal to the singular plane. Let us take note that $k_y$ and $k_z$ are constants of motion (conjugate momenta) associated to
the cyclic coordinates $y$ and $z$, then $\tilde{k}$ is a constant as well and the dispersion relation $\mathrm{H}_{am}=0$, leads to
\begin{equation} 
k_x(x)=\pm (x^2-\Tilde{k})^{1/2}\,. 
\label{elkog}
\end{equation}
The curves in Fig. \ref{fig:geodesics} correspond to planar trajectories ($k_z=0$) with values of $y_0$ and $k_y$ varying according to the arrows shown in the picture. In particular, $k_y$ takes the values 1, 2, 3, and 4. 
We see that null geodesics are complete, for their affine parameters can take values in the reals. Actually, this same property is shared by timelike geodesics, so the spacetime is timelike and null geodesically complete. In this sense, the strong curvature, naked singularity at $x=0$ is mild enough so as to prevent causal geodesic incompleteness, mainly due to its repulsive character. However, as shown in \citep[][]{maestria1}, the spacetime is not b-complete, in the sense defined by Schmidt \cite[][]{Schmidt}, see also \cite[][]{HE}. This means that some causal (non geodesic) curves having finite generalized affine parameter length, reach the singularity. In the next two sections we will study, first, what is the effect of the repulsive singularity on electrostatic fields, and later, on waves not subjected to the GO approximation, i.e. governed by the full equations (\ref{eq:rot_E_cuasi_planas}) and (\ref{eq:rot_H_cuasi_planas}). 

\section{Electrostatics}

As a simple exercise, we turn our attention first to the electrostatic case. Ignoring the time-varying terms in Faraday's and Ampère-Maxwell's laws, we have that $\bar{E}$ and $\bar{H}$ can be written in terms of scalar potentials $\phi_{E}$ and $\phi_{H}$ in the usual way, i.e., $\bar{E}=-\bar{\nabla}\phi_{E}$ and $\bar{H}=-\bar{\nabla}\phi_{H}$. These can be combined with the constitutive equations given in Eqs. (\ref{pt_1}) and (\ref{pt_2}) and the two divergence equations $\bar{\nabla} \cdot \bar{D}=0$ and $\bar{\nabla} \cdot \bar{B}=0$, to get
\begin{eqnarray}
    &&\bar{\nabla} \cdot (\textbf{K}\bar{\nabla}\phi_{E})+ (\bar{\nabla} \times\bar{\Gamma})\cdot\bar{\nabla}\phi_{H}=0\, ,\notag\\ 
     &&\bar{\nabla} \cdot (\textbf{K}\bar{\nabla}\phi_{H})- (\bar{\nabla} \times\bar{\Gamma})\cdot\bar{\nabla}\phi_{E}=0\,. \label{estaticac}
\end{eqnarray}  
Notice that $\bar{\Gamma}$ couples electric and magnetic fields, even in the static case. Taking $\bar{H}=0$, if we evaluate these equations considering (\ref{eq:K}), and writing $\phi_{E}\equiv\phi(x,y,z)$, we obtain
\begin{align}
    &\nabla^2\phi=- \frac{1}{x}\frac{\partial \phi}{\partial x}\,.\label{eq:dif_potencial}
\end{align}
The ansatz $\phi=\mathcal{X}(x)\mathcal{Y}(y)\mathcal{Z}(z)$ transforms Eq. (\ref{eq:dif_potencial}) into
\begin{equation}
x\,\mathcal{X}''+\mathcal{X}'+a^2\,x\,\mathcal{X}=0\,,\quad
    \mathcal{Y}''+b^2 \mathcal{Y}=0\,,\quad
    \mathcal{Z}''-c^2\mathcal{Z}=0\,,\notag
\end{equation}
where $a$ and $b$ are separation constants, and $c^2=a^2+b^2$. The corresponding set of solutions is
\begin{align}
    &\mathcal{X}_{a}(x)=\left\{\begin{array}{lcc}
    A_0+\tilde{A}_0\,\ln(x) & \mbox{if } & a=0\,,\\
    A_a \,J_0(a\, x)+ \tilde{A}_a \,Y_0(a\, x) & \mbox{if } & a\neq0\,,\end{array}\right.\label{compX}\\
    &\mathcal{Y}_b(y)=\left\{\begin{array}{lcc}
    B_0+\tilde{B}_0\,y & \mbox{if } & b=0\,,\\
B_b\,\operatorname{e}^{i\,b\,y}+\tilde{B}_b\,\operatorname{e}^{-i\,b\,y} & \mbox{if } & b\neq0\,,
    \end{array}\right. \\
    &\mathcal{Z}_{c}(z)=\left\{\begin{array}{lcc}
        C_0+\tilde{C}_0\,z & \mbox{if } & c=0\,, \\
C_c\,\operatorname{e}^{c\,z}+\tilde{C}_c\,\operatorname{e}^{-c\,z} & \mbox{if } & c\neq0\,,
    \end{array}\right.
\end{align}
where $A_n,B_n,C_n$ and $\tilde{A}_n,\tilde{B}_n,\tilde{C}_n$, $n=0,1,...$, are integration constants which will be ultimately determined by boundary conditions. Moreover $J_{0}(ax)$ and $Y_{0}(ax)$ are Bessel functions of the first and second kind, respectively. The final form of the potential is thus a combination of modes
\begin{equation}\label{eq:phi_general}
    \phi(x,y,z)=\sum_{a,b}  \mathcal{X}_a(x) \mathcal{Y}_b (y)\mathcal{Z}_{c}(z)\,.
\end{equation}
It is clear that, in general, divergences in the electric field will occur at $x=0$ due to the presence of $\ln(x)$ and $Y_0(a\, x)$ in (\ref{compX}). However, let us take into account that the electrostatic energy-density reads $U(x,y,z)\propto\bar{D}\cdot\bar{E}=\textbf{K}\,\bar{E}\cdot\bar{E}= |x||\bar{E}|^2$, so the actual blow up of the energy will happen if $|\bar{E}|^2=|\bar{\nabla}\phi|^2$ goes to infinity faster than $x^{-1}$ as $x\rightarrow0$. This, of course, will necessarily occur if either $\tilde{A}_0$ or $\tilde{A}_0$ (or both) are non-null. It is worth of mention the fact that such uncontrolled electrostatic energy as $x\rightarrow0$ is totally expected in view of the singular character of the spacetime \textit{itself}. 

What really is more enigmatic, is the possibility of having regular solutions as the singularity is approached. In order to illustrate this behavior, let us guarantee the regularity of the potential by taking both $\tilde{A}_{0}$ and all the $\tilde{A}_a$ null. Additionally, for the sake of conciseness, let us work in a $2D$ medium obtained from the $3D$ metric by taking constant $z$ surfaces, and let us consider the boundary condition $\phi(x_0,y)=\phi(-x_0,y)$, for arbitrary $x_0$. This condition changes the original $\mathbb{R}^2$ topology of the medium into $\mathbb{R}\times \mathbb{S}^1$, obtained by identifying the points $x_0$ and $-x_0$. This implies that $\mathcal{X}_a(x_{0})=\mathcal{X}_a(-x_{0})$ and $\mathcal{X}'_a(x_{0})=\mathcal{X}'_a(- x_{0})$. The first condition is satisfied for any $x_{0}$ since $J_0$ is an even function. As a result, its derivative is odd, and thus the boundary condition on the derivative implies $\mathcal{X}'_a(x_{0})=\mathcal{X}'_a(-x_{0})=0$. Hence, $a^2\geq0$. In particular, since $J_{0}'(a x)=-J_{1}(a x)$, we have $a\equiv a_n=j_{1,n}/x_{0}$, where $j_{1,n}$ is the $n$th zero of $J_1$. Since $J_0(ax)$ is an even function, it is sufficient to consider $n>0$. The potential (\ref{eq:phi_general}) then reads
\begin{align}
\phi(x,y)=B_0\,y+\sum_{n=1}^{\infty} A_{a_{n}}\,J_0(a_nx)\left(\operatorname{e}^{-a_n\,y}+B_{a_{n}}\,\operatorname{e}^{a_n\,y}\right)\,.
\label{eq:potencial_electrico_2_dimensiones}
\end{align}
\begin{figure}
    \centering
    \includegraphics[width=0.8\linewidth]{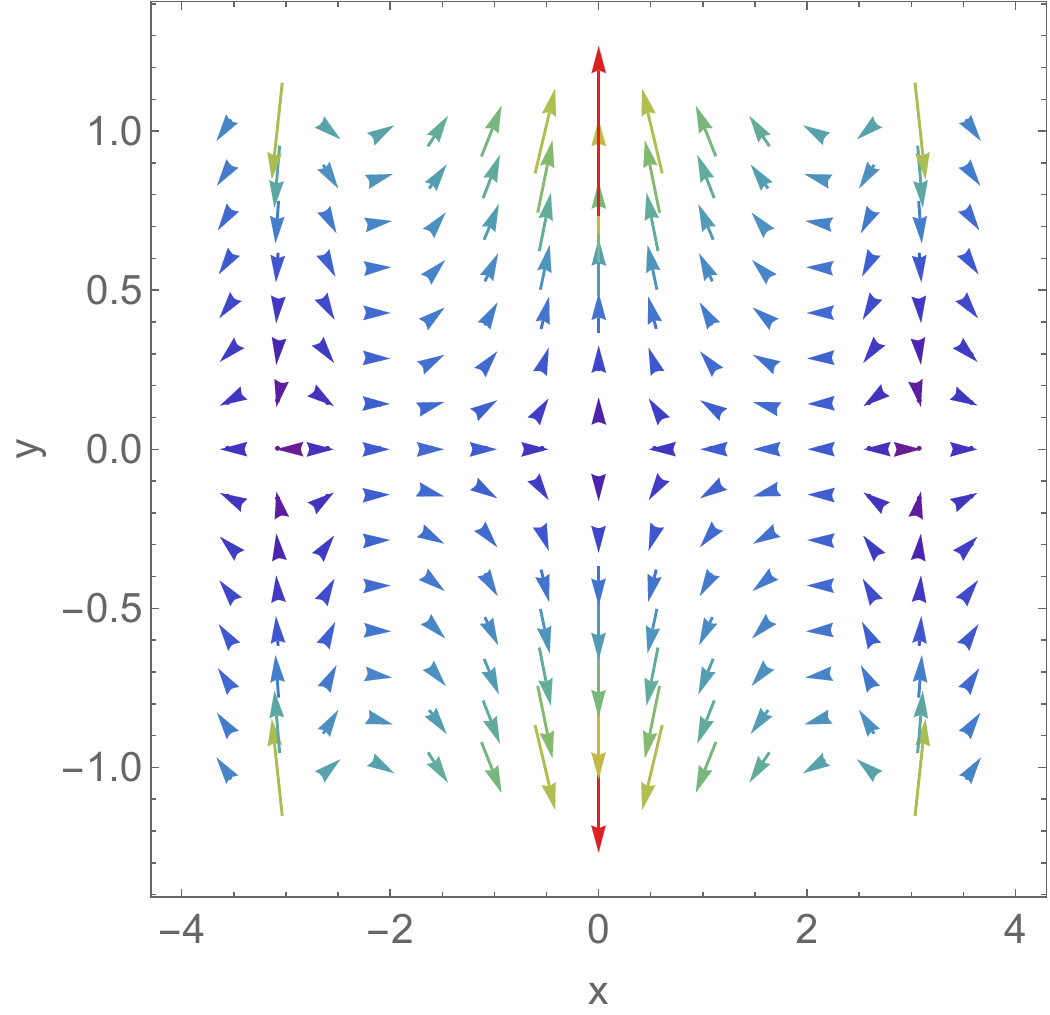}
    \caption{Vector plot of the electrostatic field in a two-dimensional case. Increasing values from blue to red.}
    \label{fig:electrostatic_field}
\end{figure}
The complete determination of the potential (up to an additive constant), will involve further boundary conditions in the $y$-direction, which can be worked out using the orthonormality relations associated to the Bessel $J_0$ function (see \cite{maestria1} for a detailed explanation). In particular, if $B_0=0$ and $B_{a_{n}}=1$ for $n\geq1$, we see that the electric field as $x\rightarrow0 $ is
\begin{equation}
\lim_{{x\to 0}}\bar{E}(x,y)\propto\sum_{n=1}^{\infty} j_{1,n}\,A_{a_{n}}\,\sinh\left({\frac{j_{1,n}}{x_{0}}\,y}\right)\,\hat{y}\,, 
\end{equation}
which is totally regular "there". Figs. \ref{fig:electrostatic_field} and \ref{fig:electrostatic_energy} display the electric field and the corresponding electrostatic energy density, respectively, for a representative example of this type.
\begin{figure}
    \centering
    \includegraphics[width=0.8\linewidth]{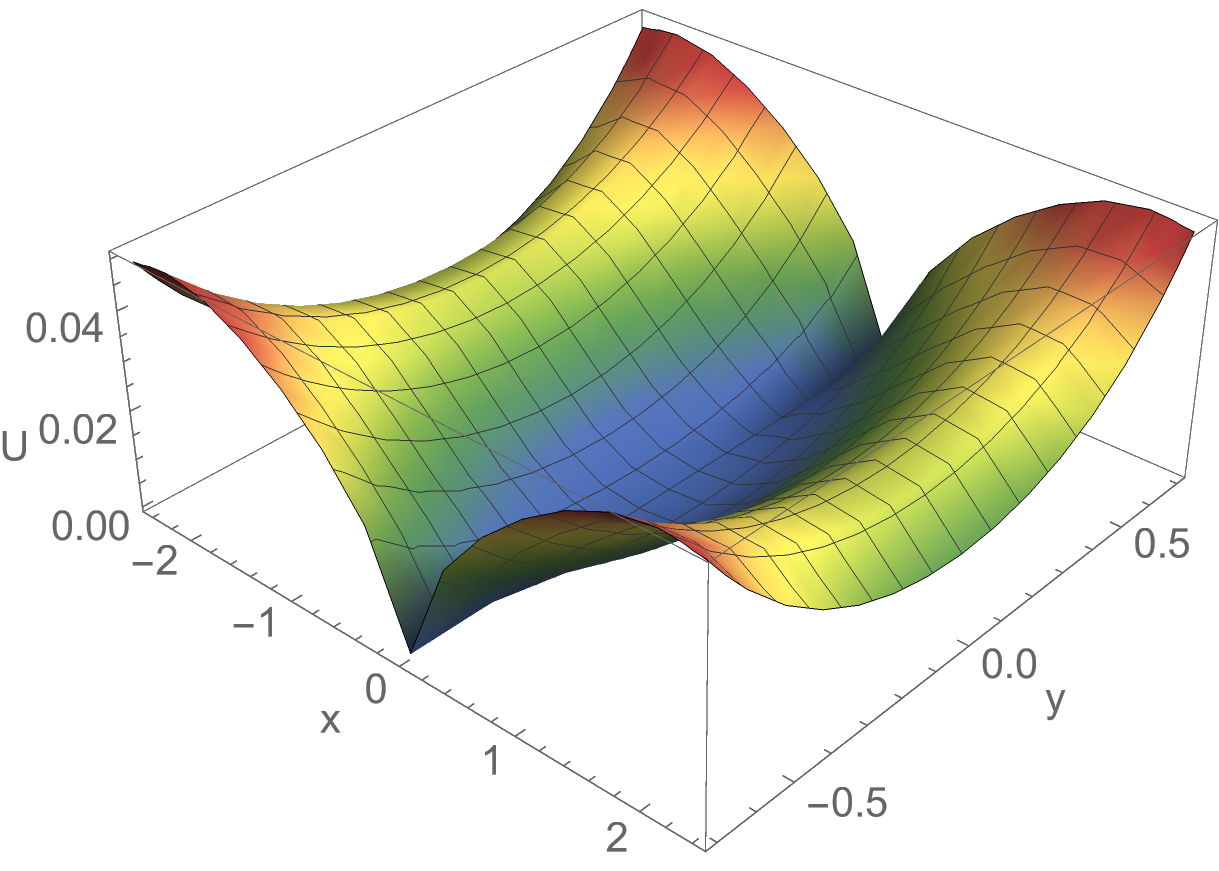}
    \caption{Electrostatic energy density in the same case as in Fig. \ref{fig:electrostatic_field}. Increasing values from blue to red.}
    \label{fig:electrostatic_energy}
\end{figure}

\section{Waves approaching the singularity}
We are now interested in the behavior of EM waves beyond the GO approximation, as they propagate toward the singular plane $x=0$. For full details, we refer the reader to \cite{maestria2}.

Let us consider an electric field of the form $\bar{E}(\bar{x})=E^z(x,y)\,\hat{z}$, and then look for planar solutions with the wave vector $\bar{k}$ associated to GO, see Eq. (\ref{elkog}). These two ansatze greatly simplify the analysis of Eqs. (\ref{eq:rot_E_cuasi_planas}) and (\ref{eq:rot_H_cuasi_planas}), whose $x$ and $y$ components are now identically zero. The $z$-component is separable by writing $E^z(x,y)=X(x)\,Y(y)$. Introducing a separation constant $c$, the final equations are
\begin{align}
    Y''&=k_0\,c^2\, Y-2\,i\,k_0\, k_y\,Y'\,,\label{eq:Y_y_cte_0}\\
    \frac{X''}{k_0}&=\left[i\left(\frac{k_x}{x}-\frac{2\,x}{k_x}+\frac{x^3}{{k_x}^3}\right)+2\,k_0\,x^2+\frac{k_0x^4}{{k_x}^2}-c^2\right]X \nonumber\\
    &-i\left(\frac{i}{k_0 x}+\frac{2\,x^2}{k_x}+2\,k_x\right)X'\,.\label{eq:X_x_cte_0}
\end{align}
We analyze separately the two cases defined by the possible values of the separation constant.

\bigskip

\begin{figure}
    \centering
    \includegraphics[width=0.9\linewidth]{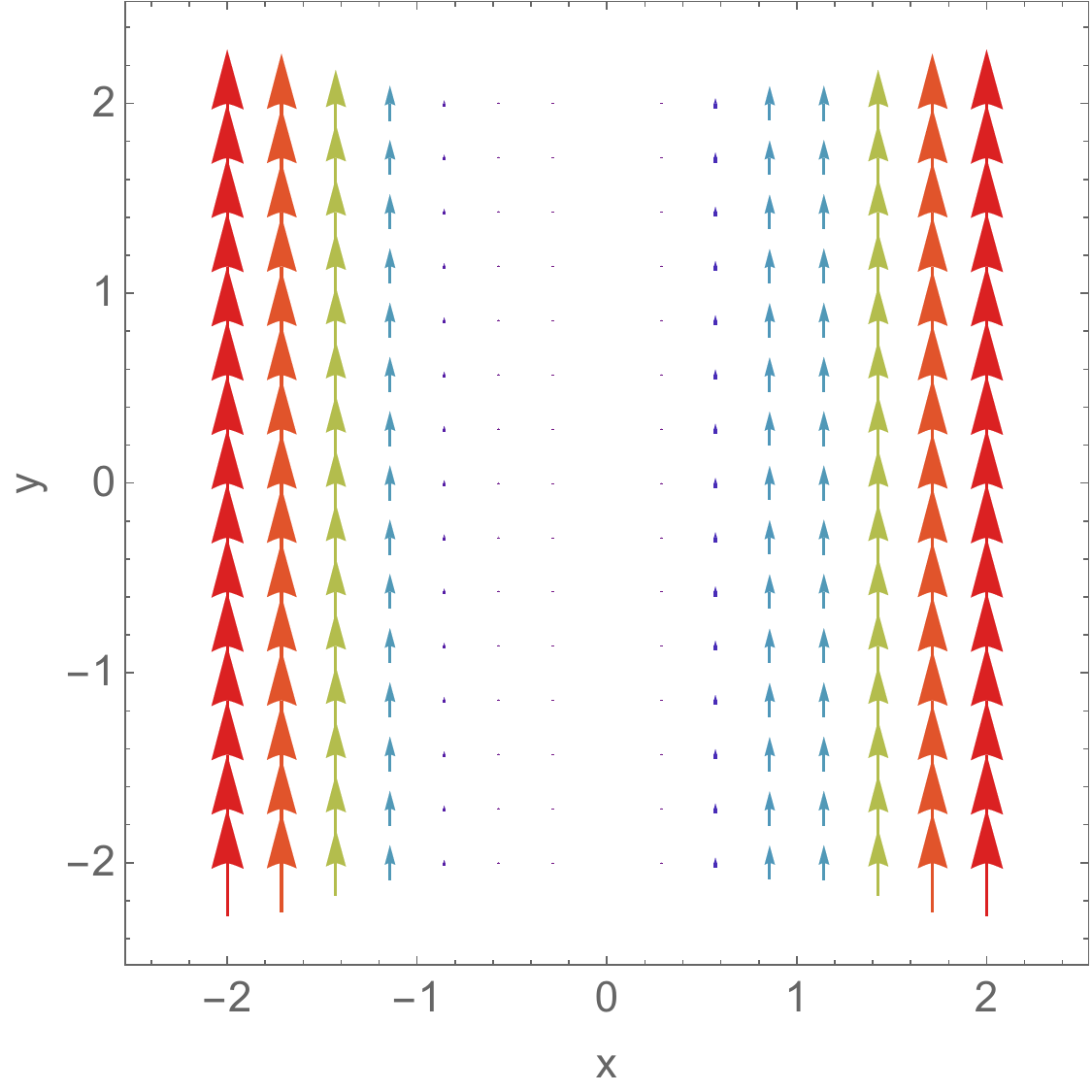}
    \caption{Time-averaged Poynting vector for one case with $c^2+{k_y}^2\neq0$. Increasing values from blue to red.}
    \label{fig:poynting_1}
\end{figure}

\textbf{(a) Case $\mathbf{c^2+k_y^2\neq 0.}$}
In this case, the solutions of Eqs. (\ref{eq:Y_y_cte_0}) and (\ref{eq:X_x_cte_0}) are
\begin{align}
    &Y=\,c_1\,\operatorname{e}^{-i\, k^{+}\, y}+c_2\,\operatorname{e}^{-i\,k^{-}\,y},\label{partey} \\
    &X=\operatorname{e}^{\frac{i\,k_0}{2}\,x\left(x-2k_{x}\right)} 
    \Big[c_3\,U\left(A,0,B\right) 
    \left.+c_4\,L_{(-A)}^{(-1)}(B)\right],\label{partex}
\end{align}
where $k^{\pm}\doteq k_0[\, k_{y}\pm({k_y}^2-c^2)^{1/2}]$, $A=i\,k_0\,\left({k_y}^2-c^2\right)/4$, $B=B(x)= -i\,k_0\, x^2$, and $c_{i}$ $i=1,2,3,4$ are integration constants in the reals. In addition, $U(a,b,z)$ is the confluent hypergeometric function of the second kind, and $L_{(\beta)}^{(\alpha)}(x)$ are the generalized Laguerre polynomials. On the other hand, using Eq. (\ref{eq:rot_E_cuasi_planas}), the non-vanishing components of the magnetic field are
\begin{align}
&H^x =X\,|x|^{-1}\left[k_y Y-i\,Y^{\prime}/k_0\right]\,,\label{hx2}\\
&H^y=-Y\,|x|^{-1}\left[\left(k_{x}+{k_{x}}^{-1}x^2\right)X-i\, X^{\prime}/k_0\right]\,.  \label{hy2}
\end{align}
It is worth noting that the hypergeometric function $U$ is finite and nonzero at $x=0$, whereas the generalized Laguerre polynomial $L_{(-A)}^{(-1)}$ vanishes at that point. Therefore, the remarkable thing here is that, for $c_3=0$, both the electric field and the magnetic field are regular (null) at $x=0$. Moreover, the time-averaged power density given by the Poynting vector, in this case vanishes at $x=0$. Actually, in a period $T$ we have
\begin{align}
    \langle \bar{S} \rangle &=T^{-1}\int_0^T \bar{S}(t)\,dt= \operatorname{Re}\left[\bar{\mathcal{E}}\times\bar{\mathcal{H}}^{*} \right]/2\nonumber \\
       &= \operatorname{Re}\left[-E^z \left(H^y \right)^{*} \hat{x} +E^z \left(H^x\right)^*\hat{y}\right]/2\label{els1},
  \end{align}
which turns, as $x\rightarrow 0$, into
\begin{align} 
     \langle \bar{S} \rangle  \approx \hat{y}\,{c_{4}}^2\,k_0 |x|^3 \operatorname{Re}(\kappa) (c_2^2 - c_1^2)/2, \label{els1f_imky0}
\end{align}
where $\kappa=k_{0}({k_{y}}^2-c^2)^{1/2}$, since $\kappa$ is either real or pure imaginary. Notice that, provided $\kappa\in\mathbb{R}$ and
$c_1 \neq c_2$, there is a net power flux always directed along $\hat{y}$ (upwards or downwards depending on the sign of $c_2^2 - c_1^2$). In Fig. \ref{fig:poynting_1}, we show one case where the flux is directed upwards. Notice that this structure for the Poynting vector is in full correspondence with the wave vector corresponding to the null geodesics (or light rays) shown in Fig. \ref{fig:geodesics}. There, the light rays are coming, let us say, from the bottom to the top of the figure, and they are reflected by the singularity. Now we have a fuller description of that, because the field amplitudes (which are not an outcome of the GO approximation) are producing a Poynting vector which is null in the whole plane $x=0$. Because of this, we have a very descriptive picture of the singularity behaving as a perfect conducting plane for \textit{any} wave number $k_{0}$, i.e, for \textit{any} frequency $\omega$. As a matter of fact, we have $\lim_{{x\to 0}}E^z(x,y)=\lim_{{x\to 0}}H^x(x,y)=0$ $\forall y$, so the perfect conductor boundary conditions are fulfilled there. On the other hand, when $c_1=c_2$, the power flux is zero because the solution for the function $Y$ consists on two contributions with the same amplitude but with opposite phases; see Eq. (\ref{partey}). 
  
\bigskip
\textbf{(b) Case $\mathbf{c=k_y=0}$.} This situation is simpler and more enlightening, since we now have a wave directed either into or away from the singularity with wave vector $\bar{k}=\pm |x|\hat{x}$. Eqs. (\ref{eq:Y_y_cte_0}) and (\ref{eq:X_x_cte_0}) acquire the form
\begin{align}
    Y''=&\,0\,,\\
   X''=& 3\,{k_0}^2\,x^2 X-i\left(i\,x^{-1}\pm4k_{0}|x|\right)X'\,,\label{dosramas}
\end{align}
which lead to 
\begin{align}
    &Y(y)=c_1\,y+c_2 \label{laformadey}\\
    &X_1(x)=\operatorname{e}^{-\frac{3}{2}i\,k_0\,x^2}   \left(c_3\operatorname{e}^{i\,k_0\,x^2}+c_4\right), \\
    &X_2(x)=\operatorname{e}^{\frac{3}{2}i\,k_0\,x^2} 
    \left(c_3\operatorname{e}^{-i\,k_0\,x^2}+c_4\right),
\end{align}
where $X_1(x)$ and $X_2(x)$ are the two possible forms of $X(x)$, due to the $\pm$ sign in (\ref{dosramas}),  according to the four schemes
\begin{enumerate}
    \item $\,\, X_1(x)$, $\forall x$ ($\bar{k}=x\,\hat{x}$),
    \item $\,\, X_2(x)$, $\forall x$ ($\bar{k}=-x\,\hat{x}$),
    \item $\,\, X_1(x)$, $x\geq0$,\, \text{and}\, $X_2(x)$, $x\leq0$ ($\bar{k}=|x|\hat{x}, \,\,\forall x$),
    \item $\,\, X_2(x)$, $x\geq0$,\, \text{and}\, $X_1(x)$, $x\leq0$ ($\bar{k}=-|x|\hat{x}, \,\,\forall x$).
\end{enumerate}
The constants $c_i$, $i:1,2,3,4$ are real numbers and they are not related to the ones of the previous case, Eqs. (\ref{partey}) and (\ref{partex}). In particular, we will conveniently take $c_2^{\,\,2}=2$, without loss of generality.

Again, the magnetic field components are obtained from  Eq. (\ref{eq:rot_E_cuasi_planas}), and they are
\begin{align}
    H^x&=-i\, c_1\,{k_0}^{-1}\,|x|^{-1} X\label{magxk1},\\
   H^y&=Y \left[\mp\,2\,X +i\,{k_0}^{-1}\,|x|^{-1}X'  \right] \label{magxk2} \,.
\end{align}
In all cases, the electric field $E^z= X(x) Y(y)$ remains regular and nonvanishing at $x=0$. In addition, setting $c_{1}=0$, ensures that the magnetic field is likewise regular at $x=0$, so that the system supports a TEM wave with finite amplitude as $x\rightarrow0 $, transporting a time-averaged power density (see Eq. (\ref{els1}))
\begin{align}
    \langle \bar{S} \rangle=-\hat{x} \operatorname{Re}\left[E^z \left(H^y \right)^{*} \right]/2. 
 \end{align}     
This can be computed right away taking into account the schemes outlined in the points (1)-(4) above, namely:

\bigskip

\textbf{Scheme 1}
\begin{align}\label{outin}
\langle \bar{S} \rangle_{out/in}&=\hat{x}\left\{\begin{array}{rl} {c_3}^2-{c_4}^2\,,&x>0 \\
      {c_4}^2-{c_3}^2\,,&x<0\,, 
      \end{array}
      \right.
\end{align}
 
 \textbf{Scheme 2 (- Scheme 1)}
\begin{align}\label{outin2}
\langle \bar{S} \rangle_{out/in}&=\hat{x}\left\{\begin{array}{rl} {c_4}^2-{c_3}^2\,,&x>0 \\
      {c_3}^2-{c_4}^2\,,&x<0\,, 
      \end{array}
      \right.
\end{align}

\textbf{Scheme 3}
\begin{equation}\label{cross1}
\langle \bar{S} \rangle_{cross}=\hat{x}\,
      (c_3^{\,\,2}-c_4^{\,\,2})\,,\forall x, 
\end{equation}

\textbf{Scheme 4 (- Scheme 3)}
\begin{equation}\label{cross2}
\langle \bar{S} \rangle_{cross}=\hat{x}\,
      (c_4^{\,\,2}-c_3^{\,\,2})\,,\forall x. 
\end{equation}

\begin{figure}
    \centering
    \begin{subfigure}{.45\linewidth}
        \includegraphics[width=\linewidth]{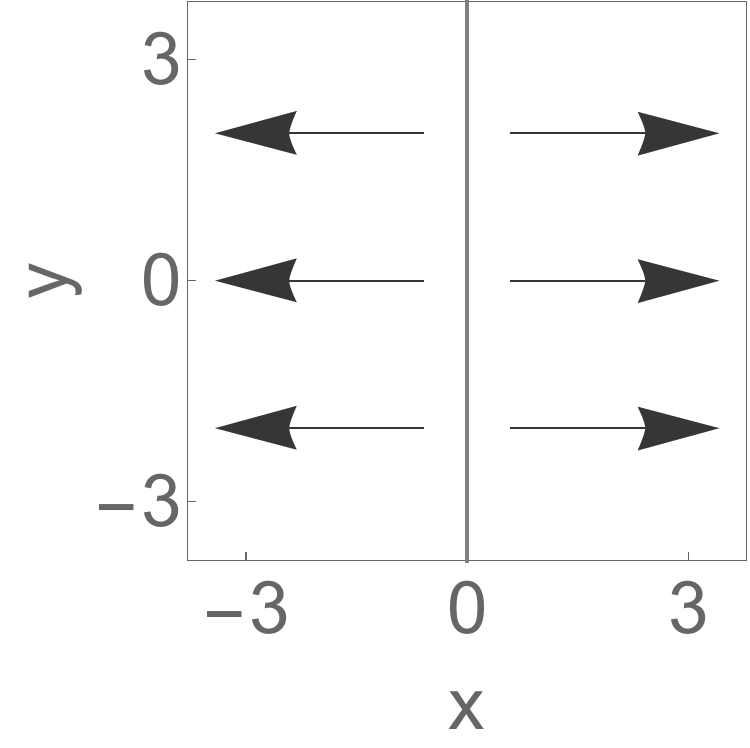}
        \caption{$S_{out}$}\label{subfig:s_out}
    \end{subfigure}
    \begin{subfigure}{.45\linewidth}
        \includegraphics[width=\linewidth]{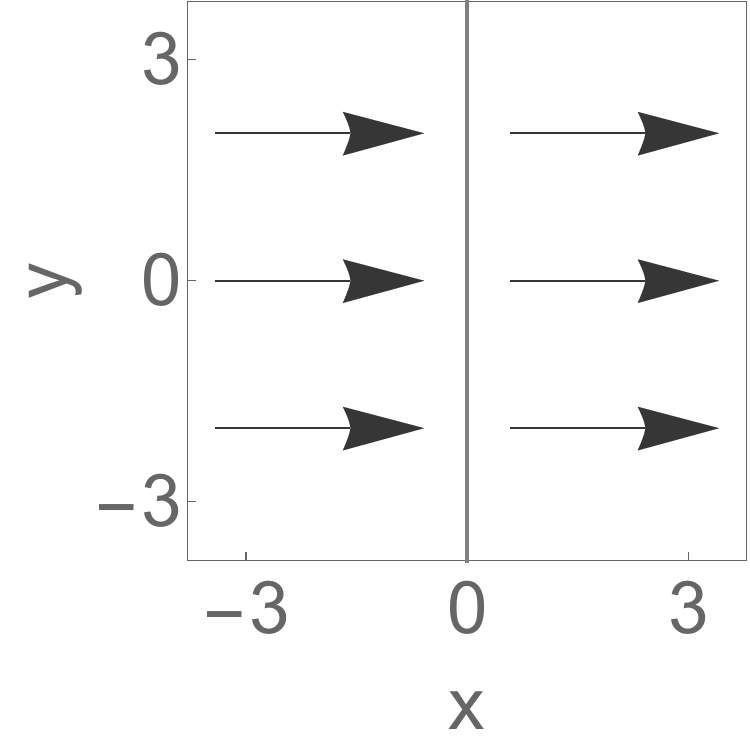}
        \caption{$S_{cross}$}\label{subfig:s_cross}
    \end{subfigure}
    \caption{Time-averaged Poynting vector for cases with $c^2+{k_y}^2=0$. (\subref{subfig:s_out}) Outgoing flux coming out from the singularity according to Scheme 1 (if ${c_3}^2> {c_4}^2$), or Scheme 2 (if ${c_3}^2< {c_4}^2$). (\subref{subfig:s_cross}) Crossing flux from left to right according to Scheme 3 (if ${c_3}^2> {c_4}^2$), or Scheme 4 (if ${c_3}^2< {c_4}^2$).}
    \label{fig:poynting_2}
\end{figure}

So, for ${c_3}^2\neq {c_4}^2$, Schemes 1 and 2 result in a discontinuous power flux, either ingoing (in) or outgoing (out), suggesting absorption or emission of waves at the singularity, respectively. On the other hand, Schemes 3 and 4 yield a constant flux as $x \rightarrow 0^{\pm}$, thus, establishing a sort of causal connection between both sides of the singularity. We illustrate these facts in Fig. \ref{fig:poynting_2}.

\section{Final Remarks}
In this contribution we have applied the Plebanski-Tamm approach to a toy model metric containing a strong curvature naked singularity and, because of the exact nature of the solutions we found, we were able to analyze the behavior of electromagnetic waves beyond the GO approximation in the vicinity of the singularity. Remarkably, we demonstrated the existence of solutions whose electromagnetic fields remain regular as $x\rightarrow 0$, both in the electrostatic and electrodynamic regimes. In particular, in the electrodynamic case, we found solutions that allow for the transmission of power flux from one side of the singularity to the other, suggesting that, within this framework, signal propagation through the singularity may be possible. In this regard, it could be genuinely argued that many of the equations we started from --as, for instance, (\ref{eq:dif_potencial}) and (\ref{eq:X_x_cte_0})-- are simply not valid at $x=0$. Let us suppose for a moment that we are not aware of the presence of a curvature singularity. We just have a differential equation like $\nabla^2\phi(x,y,z)=- x^{-1}\phi(x,y,z)_{,x}$, as our Eq.  (\ref{eq:dif_potencial}). This equation is actually not that different from some appearing in ``standard'' wave propagation in flat spacetime, as in a cylindrical waveguide, where $x$ plays the role of a radial coordinate and, of course, no real singularity exists. Clearly, the equation blows up at $x=0$. However, some solutions (in some sense, not many of them), seem to cancel that divergence; trivially, the function $\phi(x,y,z)= ay+bz$, with constant $a,b$, solves the equation, then by all means ``ignoring'' the singular point $x=0$. Are we forced to discard this perfectly regular solution because we know that the singularity is really there? Of course, in the model spacetime at hand we know that the curvature blows at $x=0$, because some curvature invariants tell us so, but still, some solutions manage to disregard that fact and behave regularly as $x\rightarrow 0^{\pm}$. In any event, the results here presented, and the conclusions derived thereof, are perfectly valid in that limit. In this particular case, it then seems that discussions about physics at $x=0$ would be more appropriate in the realm of semantics.

Although the background considered here does not correspond to a physically realistic metric, in the sense that it is not a solution of Einstein equations with any reasonable matter content, the results obtained are nevertheless significant. They indicate that electromagnetic fields may remain bounded even in the presence of a strong curvature singularity (at least in the case of a repellent naked singularity) and, moreover, that the information could in principle propagate across it. Future work may explore whether similar features arise in physically motivated spacetimes.

\section*{Acknowledgements}
FF and SMH are members of \textit{Carrera del Investigador Cient\'{i}fico} (CONICET). Their work is supported by CONICET and Instituto Balseiro (UNCUYO). JMP is a PhD student supported by CONICET.

\end{document}